\documentclass[twocolumn,showpacs,amsmath,amstex,amssymb,mathfonts,prl]{revtex4-1}
\usepackage{amsthm,amsfonts,graphicx,verbatim}

\usepackage{dcolumn}   
\usepackage{verbatim}
\usepackage{tipa}
\usepackage{wasysym}
\usepackage{esint}
\usepackage{color}

\usepackage{amsmath}
\usepackage{amssymb}
\usepackage{amsthm}
\usepackage{amsfonts}
\usepackage{listings}
\usepackage{enumerate}
\usepackage{latexsym}
\usepackage{psfrag}
\usepackage{bm}
\usepackage{subfigure}

\begin{document}

\title{Model Wavefunctions for the Collective Modes and the Magneto-roton Theory of the Fractional Quantum Hall Effect}

\author{Bo Yang$^1$, Zi-Xiang Hu$^{2,3}$, Z. Papi\'c$^{2}$, and F. D. M. Haldane$^1$}
\affiliation{$^1$ Department of Physics, Princeton University, Princeton, NJ 08544}
\affiliation{$^2$ Department of Electrical Engineering, Princeton University, Princeton, NJ 08544}
\affiliation{$^3$ Department of Physics, ChongQing University, ChongQing, 440004, China}


\date{\today}
\begin{abstract}
We construct model wavefunctions for the collective modes of fractional quantum Hall (FQH) systems. The wavefunctions are expressed in terms of symmetric polynomials characterized by a root partition that defines a ``squeezed" basis, and show excellent agreement with exact diagonalization results for finite systems. In the long wavelength limit, we prove the model wavefunctions are identical to those predicted by the single-mode approximation, leading to intriguing interpretations of the collective modes from the perspective of the ground state guiding center metric. 
\end{abstract}

\maketitle

One of the main driving forces in characterizing the topological phases arising in the context of the fractional quantum Hall (FQH) effect~\cite{tsg} has been the construction of the model wavefunctions 
and their parent Hamiltonians~\cite{laughlin, prange, mr, read_rezayi}. Depending on the nature of the effective interaction within a partially filled Landau level, numerous phases appear, 
ranging from crystals to incompressible liquids with fractional~\cite{laughlin}, and possibly non-Abelian~\cite{mr} quasiparticle statistics. Within the lowest Landau level (LLL), the kinetic energy of the electrons is a constant, and all dynamics comes from the Coulomb repulsion~\cite{prange}. Thus the essential features of the many-body ground states and the low-lying excitations are solely determined by the particle statistics and quantum fluctuations that minimize the repulsion.  Laughlin's wavefunction~\cite{laughlin} is a prominent example: the statistics is taken care of by the odd power in the Jastrow factor, and for rotationally invariant system, the energy is minimized by placing all zeros of the wavefunction on the electrons. Subsequently, an intuitive picture of ``composite fermions" was put forward by Jain~\cite{jain}, that lead to a classification of the numerous fractional states observed in experiment, and provided a numerical procedure for constructing various model wavefunctions~\cite{jain}. While much effort has been devoted to formulating FQH ground-state wavefunctions and those of charged excitations (quasihole and quasielectron), relatively little progress has been made in understanding the \emph{neutral} excitations since the seminal work of Girvin, MacDonald and Platzman that introduced the single mode approximation (SMA) to describe the lowest excitation in terms of a neutral density wave or the ``magneto-roton"~\cite{girvin}. Very recently, this question was revisited and explicit wavefunctions for the neutral collective excitations were proposed for the Abelian, as well as non-Abelian, FQH states utilizing a multicomponent composite fermion approach~\cite{sreejith, rodriguez}.

In this Letter we construct model wavefunctions for the collective modes of FQH states by extending the Jack polynomial description~\cite{andrei,andrei_collective}. We model the lowest neutral excitation as a dipole formed by a single quasielectron and a single quasihole. At the Laughlin $\nu=1/3$ filling, we obtain a single bosonic mode that corresponds to the magneto-roton; in the Moore-Read $\nu=5/2$ case, we obtain in addition a neutral fermionic mode~\cite{rg,nf} that stems from the non-Abelian statistics. We identify the long-wavelength limit of the bosonic modes as a ``spin-2" excitation (analogous to the graviton), and that of the fermionic mode as a ``spin-3/2" (with a possible analogy to the gravitino). Our model wavefunctions show excellent agreement with exact diagonalization results at all wavelengths. In the limit of long wavelengths, we present a proof that the model wavefunctions become equivalent to the SMA result, which thus remains an accurate description even at energies lying above the threshold of the roton-pair continuum. 

We first review some basic properties of the Jack polynomials. The Jacks are symmetric multivariable polynomials $J^\alpha_\lambda(z_1,z_2,\cdots z_{N_e})$, parametrized by a number $\alpha$ and a root partition $\lambda$ with length $l_\lambda\le N_e$, $N_e$ being the number of electrons.
For bosonic FQH states, the statistics of the system restricts us to only symmetric monomials with a fixed total degree, which corresponds to a fixed angular momentum on a disk or a sphere~\cite{haldane}. Fermionic FQH states are obtained by multiplying the bosonic counterpart with a Vandermonde determinant. 
The root partitions encode the clustering properties that effectively ``keep electrons apart" from each other; a ``$(k,r,N_e)$-admissible" partition is such that no more than $k$ electrons are found in $r$ consecutive orbitals. The root partition tells us how the wavefunction vanishes when particles are brought together. For example, the root partition of the ground state and the quasihole states at $\nu=1/m$ Laughlin filling contains no more than one electron in $m$ consecutive orbitals. The corresponding Jack vanishes at least with a power of $m$ when two electrons approach each other.

From a practical point of view, the Hilbert space we are interested in, after reduction by particle statistics and symmetry, only needs to include polynomials with appropriate clustering properties (whereby Jacks are special cases). In fact, for the ground state and single quasihole states of the Read-Rezayi series (including the Laughlin and Moore-Read states), such Hilbert space is one-dimensional and spanned by \emph{single} Jack polynomial with $\alpha=-\frac{k+1}{r-1}$ that can be obtained directly via recursive relations~\cite{productrule}. Quasielectron states, on the other hand, are more complicated~\cite{andrei_qe} because they contain local defects where electrons are forced to get closer to each other than allowed in the ground state. The same difficulty arises in collective modes which consist of quasielectron-quasihole pairs. In these cases the reduced Hibert space is no longer one-dimensional with a single Jack polynomial. 

We now proceed to construct explicit model wavefunctions for the magneto-roton mode in the Laughlin state at $\nu=1/3$ filling. We work in the spherical geometry with a monopole of strength $2S$ placed at the center~\cite{haldane}. The root partition for the Laughlin ground state is well known: $\lambda=\{100100100100\cdots 1001\}$~\cite{andrei}. The clustering property in this case means that no more than a single electron can exist in three consecutive orbitals~\cite{com}. The collective mode in the limit of momentum $\mathbf{k}\to 0$ consists of two quasielectron-quasihole pairs, forming a quadrupole. As the momentum increases, a dipole moment develops with the separation of one quasielectron-quasihole pair. This is summarized in the explicit set of root partitions as follows: 
\begin{eqnarray}\label{laughlinwfs}
&&\d{1}\d{1}0\textsubring{0}\textsubring{0}01001001001001\cdots    L=2\nonumber\\
&&\d{1}\d{1}0\textsubring{0}010\textsubring{0}01001001001\cdots    L=3\nonumber\\
&&\d{1}\d{1}0\textsubring{0}010010\textsubring{0}01001001\cdots    L=4\nonumber\\
&&\d{1}\d{1}0\textsubring{0}010010010\textsubring{0}01001\cdots    L=5\nonumber\\
&&\d{1}\d{1}0\textsubring{0}010010010010\textsubring{0}01\cdots    L=6\nonumber\\
&&\vdots
\end{eqnarray} 
We label the states by their total angular momentum $L$ on the sphere. Note the ground state with the root partition $\{1001001\ldots1001\}$ has $L=0$, and the excitation with the smallest momentum that can be created is $L=2$. In Eq.(\ref{laughlinwfs}) the black dot schematically indicates the position of a quasielectron, while the white dot that of a quasihole. To determine the position of a quasiparticle, we look at any three consecutive orbitals in the root partitions above, and count the number of electrons to see if it violates the ground state clustering property. In this particular case, if there is more (less) than one electron, we then have a quasielectron (quasihole), which is located right below the middle of the three consecutive orbitals . Due to rotational invariance on the sphere, we next impose the highest weight condition on the wavefunctions $|\psi_\lambda^L\rangle$ to single out the state with quasiparticles piled up at the north pole:
\begin{eqnarray}\label{hwt}
\nonumber  L^+|\psi^L_\lambda\rangle=0,\\
|\psi^L_\lambda\rangle=\sum_{\mu\preceq\lambda}a_\mu m_\mu
\end{eqnarray}
where $m_\mu$ are monomials with partition $\mu$~\cite{andrei}. The summation is over all partitions $\mu$ that can be squeezed from the root partition $\lambda$. For example, $m_{\{1001\}}\sim z_1^3-z_2^3,m_{\{0110\}}\sim z_1^2z_2-z_1z_2^2$, and $\{0110\}$ is squeezed from $\{1001\}$. The constraints in Eq.(\ref{hwt}) substantially reduce the Hilbert space dimension (e.g., the basis dimension is less than 20 for 10 particles). The resulting lowest-energy eigenstates of the Hamiltonian, restricted to this Hilbert space, are very good approximations to the exact magneto-roton mode.

The innovation we implement here is to impose an additional constraint that can be formally expressed as
\begin{eqnarray}\label{hwt2}
\hat{V}_1c_1c_2|\psi_\lambda^L\rangle=0.
\end{eqnarray}
Here $\hat{V}_1$ is the operator corresponding to the first Haldane pseudopotential~\cite{prange}, and $c_i$ annihilates an electron at $i$th orbital. This additional constraint renders $|\psi^L_\lambda\rangle$ unique by enforcing the following clustering property: the wavefunction is vanishing only when two or more clusters of two particles coincides in real space.

The resulting implementation is numerically much less expensive, with variational energies only slightly above the ones obtained in the Hilbert space defined by constraints (2), and improving with the increase in system size. We note that the model wavefunctions $|\psi_\lambda^L\rangle$ inherit rich algebraic structures from the underlying Jacks. When the geometric normalization factors on the sphere are removed, the coefficients of the decomposition in Fock space are integers, with the coefficient of the root configuration normalized to 1. Furthermore, they satisfy a ``product rule"~\cite{productrule,ronny} if the first five orbitals are treated as one ``big" orbital, which allows us to generate a large subset of coefficients recursively. This suggests the product rule is not restricted to pure Jacks, and is robust against local defects of the wavefunction. An approximation to $|\psi_\lambda^L\rangle$ can be built from the product rules; the overlap between the approximate and exact model wavefunctions is high and increases with system size (see Table I). These properties reduce the computational cost for generating these model wavefunctions compared to direct diagonalization.
\begin{table}\label{overlaps}
\caption{ The overlap of the approximate model wavefunctions constructed from product rules and the true model wavefunctions.}
\begin{tabular}{|l|l|l|l|l|}
\hline
No. of electrons & 9 &10 & 11 & 12 \\ \hline
L=2 & 89.83\% & 90.13\% & 90.31\% & 90.42\% \\ \hline
L=3 & 86.42\% & 86.99\% & 87.37\% & 87.63\% \\ \hline
L=4 & 83.63\% & 84.59\% & 85.23\% & 85.69\% \\ 
\hline
\end{tabular}
\end{table}

\begin{figure}
\includegraphics[width=9cm,height=7cm]{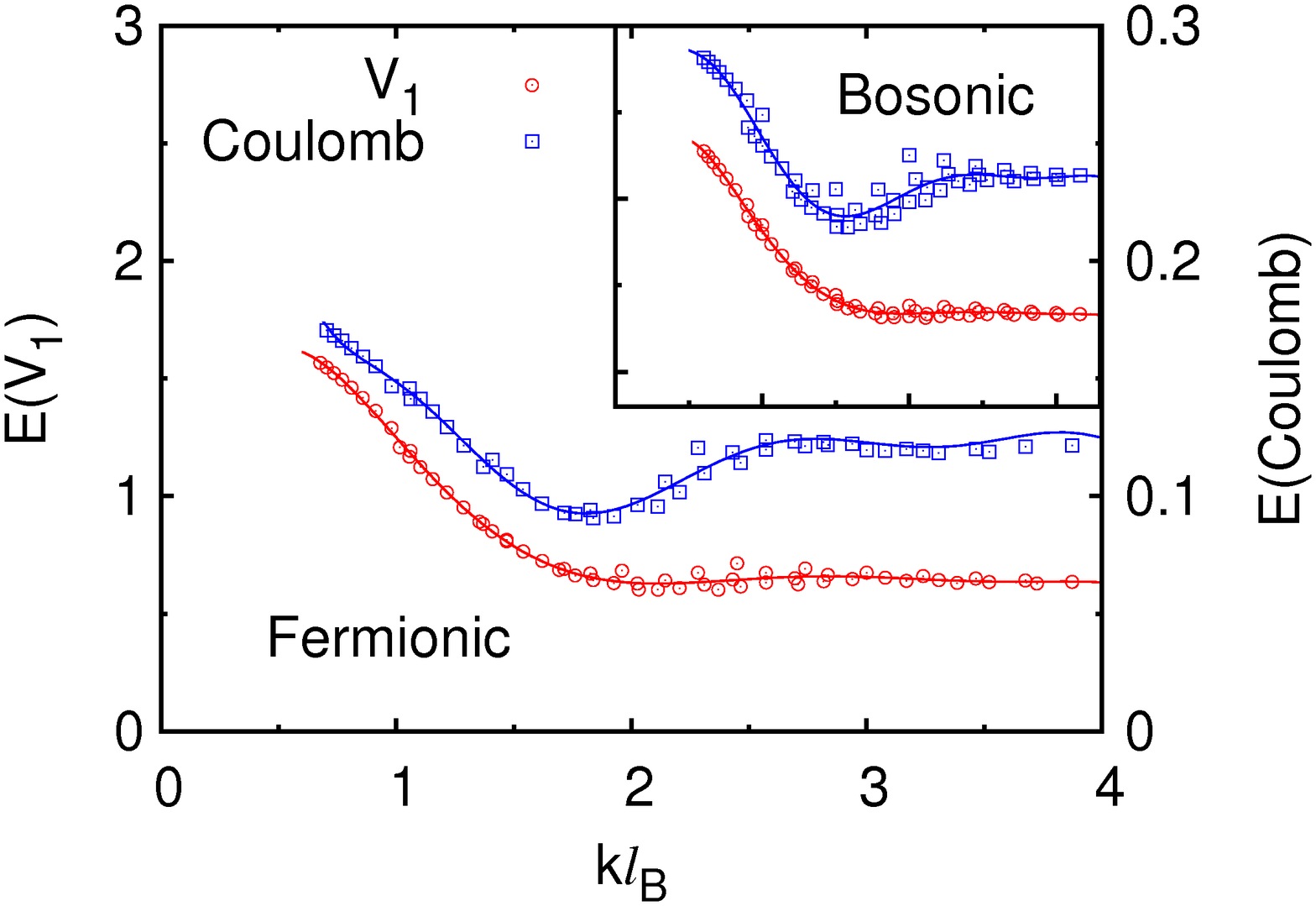}
\caption{(Color online). The variational energy of the model wavefunctions defined by Eqn (2) and (3), against $\widehat{V}_1$ (left axis, arbitrary units) and Coulomb Hamiltonian (right axis, in units of $e^2/\epsilon \ell_B$), plotted as a function of momentum. The data is generated from system sizes ranging from 6 to 12 electrons (the inset shows the same plot for the bosonic Laughlin state).}
\label{fig:laughlin}
\end{figure} 
We now proceed to evaluate the variational energies of the model wavefunctions obtained via constraints Eq.(\ref{hwt}) and (\ref{hwt2}). In Fig.~\ref{fig:laughlin}, the variational energies are plotted versus momentum $k=L/\sqrt{S}$, where $N_{\rm orb}=2S+1$ is the number of orbitals in the LLL. We include the data for a number of system sizes and rescale the magnetic length $\ell_B$ by a factor $\sqrt{S/N_{\rm orb}}$ to minimize the finite size effects. For the model $\widehat{V}_1$ Hamiltonian and Coulomb Hamiltonian, the dispersion obtained using the model wavefunction is in excellent agreement with the results from exact diagonalization, both in small $k$ and large $k$ regime. Our model wavefunctions compare favorably with the exact diagonalization eigenstates, with 99\% overlap for 10 electrons. 

\begin{figure}
\includegraphics[width=9cm,height=7cm]{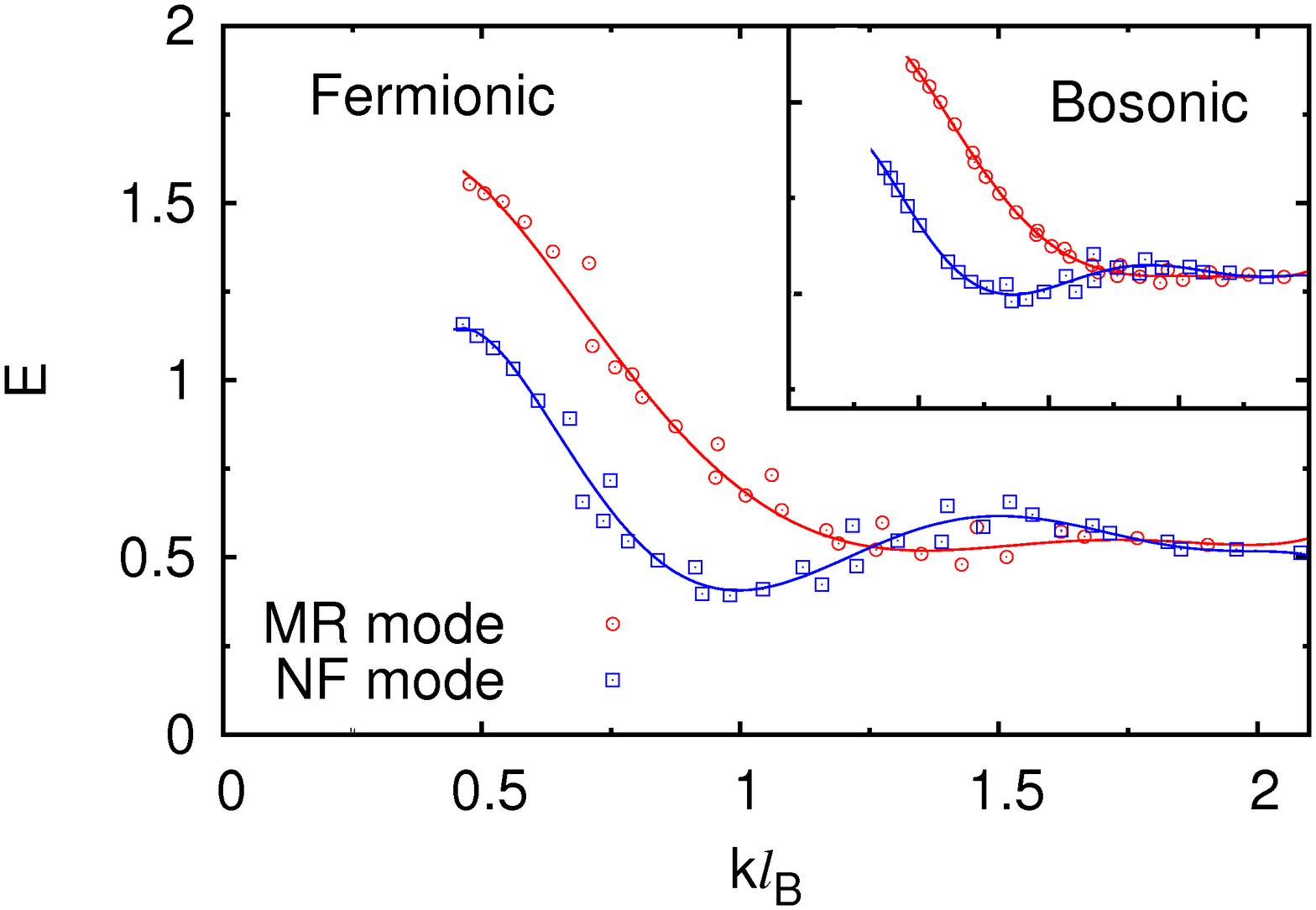}
\caption{(Color online). The variational energy of the model wavefunctions for the magneto-roton (MR) mode and the neutral fermion (NF) mode, evaluated against the 3-body Hamiltonian. The data is generated from system sizes ranging from 5 to 17 electrons, where the odd number of electrons contribute to the NF mode, and the even number of electrons contribute to the magneto-roton mode. (The inset shows the same plot for the bosonic Moore-Read state)}
\label{fig:mr}
\end{figure}

Using the same approach we can construct the collective-mode wavefunctions for the entire Read-Rezayi series. As as an example, we consider an interesting case of the Moore-Read state, where in addition to the bosonic mode, we also obtain a mode that corresponds to the unpaired electron -- the neutral fermion (NF) mode~\cite{nf}.
On the sphere, the root configurations of the two modes are given by
\begin{eqnarray}\label{mrwf}
&&11100100110011\cdots L=2,\quad111000110011\cdots L=3/2\nonumber\\
&&11100101010011\cdots L=3,\quad111001010011\cdots L=5/2\nonumber\\
&&11100101010101\cdots L=4,\quad111001010101\cdots L=7/2\nonumber\\
&&\vdots
\end{eqnarray}
The Moore-Read ground-state root partition is given by 2 electrons in 4 consecutive orbitals~\cite{andrei}. Similarly to the Laughlin state, any deviation from the uniform background density yields the position of the quasihole/quasielectron (not labeled in Eq.(\ref{mrwf})). In Eq.(\ref{mrwf}), the left column with an integer angular momentum is the magneto-roton mode. The NF mode with a half-integer angular momentum is given in the right column. Unique model wavefunctions can be constructed by imposing the constraint Eq.(\ref{hwt}), and in addition a modified constraint Eq.(\ref{hwt2}) that reads $H_{\rm 3b}c_1c_2c_3|\psi_\lambda^{\rm MR}\rangle=0$, where $H_{\rm 3b}$ is the Moore-Read three-body Hamiltonian. Their variational energies are plotted in Fig.~\ref{fig:mr}.

We would like to emphasize for both the Laughlin and Moore-Read case, the collective modes enter the multi-roton continuum in the long wavelength limit. The continuum starts at energy that is double the energy gap of the roton minimum. While this makes exact diagonalization ambiguous, the root partitions give clear physical interpretation for the modes for the entire momentum range. Interestingly, the $L=1$ state (and $L=1/2$ state for the neutral fermion mode) vanishes with the set of constraints we impose. Thus in the long wavelength limit, the collective mode is given by a quadrupole excitation. In light of the geometrical picture~\cite{haldane2} of the FQH effect, we identify the magneto-roton mode as a spin-2 ``graviton", and the NF mode as a spin-$\frac{3}{2}$ ``gravitino", or the ``supersymmetric partner" of the magneto-roton mode. 

Recently, model wavefunctions for the excitations at $\nu=1/2$ filling for bosons were also obtained in a multicomponent composite fermion picture~\cite{sreejith,rodriguez}. We have confirmed that the wavefunctions of 
Ref.~\cite{rodriguez} are numerically identical to ours in finite size systems. The approach of Ref.~\cite{rodriguez}, though completely different at the outset, arrives at the wavefunctions that satisfy 
the same clustering properties as ours, which fixes them to be unique. The uniqueness property motivates us to investigate the SMA wavefunctions obtained from the ground state $|\psi_0\rangle$ by the guiding center density modulation, $|\psi_\mathbf{k}\rangle=\hat{\rho}_\mathbf{k}|\psi_0\rangle$, where $\hat{\rho}_\mathbf{k}=\sum_i e^{i\mathbf{k}\mathbf{R}_i}$ is the guiding center density operator. 
The SMA yields excitation energies manifestly as a property of the ground state. Though it successfully predicts the magneto-roton minimum of the collective mode of FQH states at $\nu=1/m$, there is some ambiguity in the limit of $\mathbf{k}\rightarrow 0$ when the SMA variational energy enters the roton-pair continuum, eluding comparisons with exact diagonalization. It is thus useful to compare the SMA prediction with our model wavefunctions in the small $\mathbf{k}$ limit, since the latter are valid in that regime and have transparent physical properties given by the root partitions.

\begin{figure}
\includegraphics[width=9cm,height=7cm]{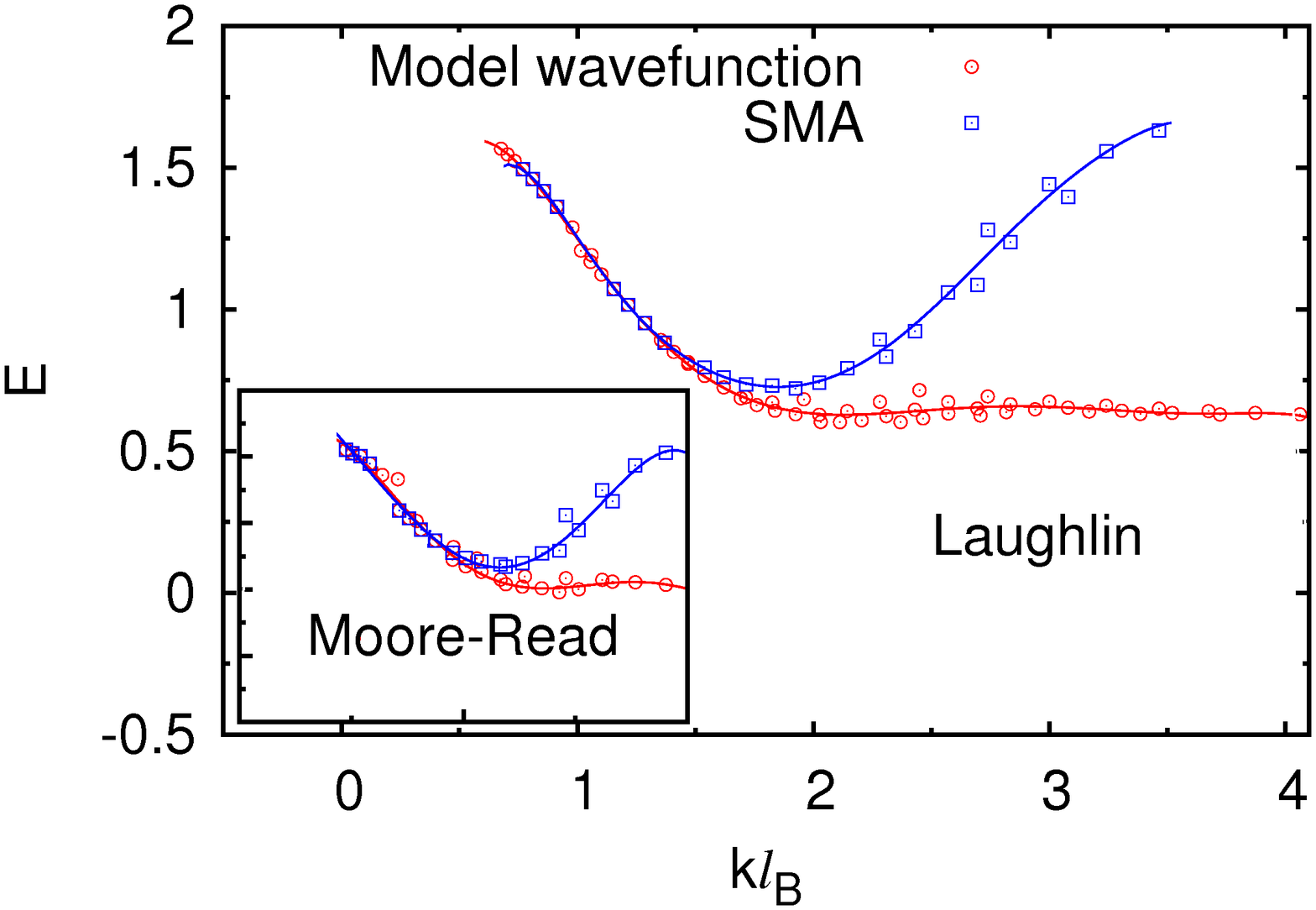}
\caption{(Color online). The variational energies for the SMA model wavefunctions compared to our model wavefunctions for Laughlin state at $\nu=1/3$ filling. (The inset shows the same comparison for the magneto-roton mode of the Moore-Read state)}
\label{fig:sma}
\end{figure}

The SMA construction can be adapted to the sphere as follows. The ground state on a sphere has the total angular momentum $L=0$, and the SMA wavefunction with total angular momentum $L$ is obtained by boosting one electron with orbital angular momentum $L$. The projection into the LLL is equivalent to the projection of the boosted single-particle state into the sub-Hilbert space of the total spin $S$. Formally we have
\begin{eqnarray}\label{smawf}
|\psi^{\rm SMA}_{LM}\rangle=\sum_i \hat{C}^{S,L,S}_{m_i+M,M,m_i}|\psi_0\rangle,
\end{eqnarray}
where $i$ is the electron index, and $\hat{C}^{S,L,S}_{m',M,m}$ is defined by its action on the single electron state $\hat{C}^{SLS}_{m'Mm}|m\rangle=C^{SLS}_{m'Mm}|m'\rangle$, where  $C^{SLS}_{m'Mm}=\langle m'|\hat{Y}^{LM}|m\rangle$ are the Clebsh-Gordon coefficients, and $\hat{Y}^{LM}$ are the spherical harmonics. This is a result of the Wigner-Eckart Theorem, and due to rotational invariance we can set $M=L$ in Eq.(\ref{smawf}). The dispersion of the SMA wavefunctions is plotted in Fig.~\ref{fig:sma} along with that of our model wavefunctions. 

For small momenta the variational energies of the two classes of wavefunctions agree very well, while the SMA mode evidently becomes invalid for momenta larger than the magneto-roton minimum. Note that at $L=2,3$ the SMA wavefunctions only involve the elements of the basis squeezed from the same root partition that defines our model wavefunctions. Taking the Laughlin $1/3$ as an example, we now prove the SMA wavefunctions are actually identical to ours at $L=2,3$. By the product rule of the Jack polynomial, we can write
\begin{eqnarray}\label{product}
 |\psi_0\rangle\sim J^\alpha_{\lambda_1}\otimes J^\alpha_{\lambda_2}+J^\alpha_{\lambda_3}\otimes J^\alpha_{\lambda_4}+|\bar{\psi}_0\rangle,
\end{eqnarray}
where we suppressed the relative coefficients because they are unimportant for the proof. The partitions $\lambda_1=\{10010\},\lambda_2=\{01001001\cdots\},\lambda_3=\{10001\},\lambda_4=\{10001001\cdots\}$, and $|\bar{\psi}_0\rangle$ involves the rest of the squeezed basis. It is easy to check that $c_1c_2\sum_i\hat{C}^{S,L,S}_{m_i+L,L,m_i}|\bar{\psi}_0\rangle=0$. We thus have
\begin{eqnarray}\label{v1product}
\hat{V}_1c_1c_2|\psi^{\rm SMA}_{LL}\rangle&\sim&\hat{V}_1\{0000\}\otimes (J^\alpha_{\lambda_2}+tJ^\alpha_{\lambda_4})=0
\end{eqnarray}
Again, the coefficients are suppressed in Eq.(\ref{v1product}), and $t=0$ for $L=2$. Thus the SMA wavefunctions satisfy exactly the same constraints as the our model wavefunctions, which makes them identical. Note that for $L>3$ the SMA wavefunctions contain unsqueezed basis components with respect to the root partitions used in our model wavefunction, and the proof breaks down. 

In conclusion, we demonstrated a numerically efficient method of constructing accurate model wavefunctions for the collective modes in FQH systems. The wavefunctions are identified with the SMA wavefunctions in the long wavelength limit. This result reveals a crucial link between the ``graviton" mode and the SMA mode at long wavelengths, which plays an important role in the geometrical theory of the FQH effect~\cite{haldane2}. For realistic Coulomb or pseudopotential interactions, the ``graviton" mode decays into multi-roton pairs and appears experimentally inaccessible. It would be interesting to see if the interaction can be tuned in such a way as to expose the ``graviton" mode at $\mathbf{k}\to 0$ below the roton-pair continuum.

{\sl Acknowledgements}. We thank B. A. Bernevig for useful discussions. This work was supported by DOE grant DE-SC$0002140$.

\bibliography{paper}

\begin{thebibliography}{99}
\bibitem{com}
The bosonic counterpart has no more than a single particle in two consecutive orbitals, yielding a Jack given by $J^{-2}_\lambda$, which is a function that vanishes as a power of 2 when two particles approach each other, i.e. the Laughlin wavefunction.
\end{thebibliography}

\end{document}